\documentclass[conference]{IEEEtran}
\IEEEoverridecommandlockouts
\usepackage{cite}
\usepackage{amsmath,amssymb,amsfonts}
\usepackage{algorithmic}
\usepackage{graphicx}
\usepackage{float} 
\usepackage{subcaption}
\usepackage{textcomp}
\usepackage{xcolor}
\def\BibTeX{{\rm B\kern-.05em{\sc i\kern-.025em b}\kern-.08em
    T\kern-.1667em\lower.7ex\hbox{E}\kern-.125emX}}
\begin{document}

\title{Polarization-diverse Detection at Microwave Frequencies Using A Passive Metasurface Aperture\\
}


\author{\IEEEauthorblockN{Md. Abrar A Mushfik\IEEEauthorrefmark{1},
Mohammad Ali Kaisar\IEEEauthorrefmark{1},
Mohiminul Islam Bhuiyan Sahed\IEEEauthorrefmark{1}, and 
Idban Alamzadeh\IEEEauthorrefmark{1}\IEEEauthorrefmark{2},~\IEEEmembership{Member,~IEEE}}
\IEEEauthorblockA{\IEEEauthorrefmark{1}Dept. of Electrical and Electronic Engineering,
International Islamic University Chittagong, Chittagong, Bangladesh}
\IEEEauthorblockA{\IEEEauthorrefmark{2}School of Electrical, Computer, and Energy Engineering, Arizona State University, AZ, USA}
\thanks{Manuscript received December 1, 2012; revised August 26, 2015. 
Corresponding author: Idban Alamzadeh (email: Idban.Alamzadeh@asu.edu).}}

\maketitle
\begin{abstract}
Metasurfaces' ability to control electromagnetic wave propagation has led to a rapid paradigm shift in wireless operation. These metasurfaces are often called reconfigurable intelligent surfaces (RISs) due to active tuning elements distributed across the meta-atoms comprising the metasurface array. However, each of these dynamic meta-atoms requires additional DC power lines and biasing circuitry for active tuning. Additionally, achieving polarization diverse operations using compact metasurface configurations is challenging due to the complexity involved in polarization detection. To address these limitations, we propose a passive metasurface array architecture that is both polarization sensitive and capable of altering radiation patterns with frequency diversity. In particular, we designed a polarization-sensitive meta-atom model with added randomness in the scattering behavior and extended it to a polarization-diverse-frequency-selective array. By capturing the electric fields scattered off from the metasurface, we can numerically acquire the polarization information of the incoming signal. The proposed polarization-diverse array can simplify the polarization measurement techniques and may find its application in polarization sensitive sensing and imaging operations.
\end{abstract}
\section{Introduction}
Metasurfaces, composed of subwavelength engineered scatterers, have emerged as a powerful platform for manipulating electromagnetic waves across microwave, terahertz, and optical frequency regimes \cite{A37,A39,B37,B38}. By tailoring the local electromagnetic response of individual meta-atoms, metasurfaces enable precise control over amplitude, phase, and polarization within electrically thin apertures\cite{A44 ,A40 ,B39 ,B40}. These properties have enabled a wide range of wavefront-engineering functionalities, including beam steering \cite{A37,A39,A44}, focusing\cite{A40}, holography\cite{A41},\cite{A42}, polarization manipulation \cite{A43},\cite{A2}, and scattering control\cite{A19},\cite{B41}, with applications spanning wireless communications, computational imaging, radar, and sensing \cite{A1 ,A5 ,A14 ,B42 ,B43}.

A major advancement in metasurface-based systems is the development of software-controlled reconfigurable intelligent surfaces (RIS). In these architectures, active tuning elements—such as PIN diodes, varactors, or switchable loads—are embedded within meta-atoms to dynamically reconfigure the scattering response of thesurface \cite{A6,A7,B44,B45}. Such reconfigurability enables real-time manipulation of reflected wavefronts and has been explored extensively for coverage enhancement, interference mitigation, and channel shaping in next-generation wireless networks \cite{A8,A10,B46,B47}.

Despite their flexibility, RIS platforms introduce significant hardware complexity due to dense biasing networks, distributed power delivery, and multilayer routing requirements\cite{A6}, \cite{A7}, \cite{B45}, \cite{B48}. As aperture size increases, these factors limit scalability, increase power consumption, and complicate fabrication\cite{A10}, \cite{B46}, \cite{B49}.
In addition to hardware overhead, actively reconfigurable metasurfaces impose substantial computational complexity. Effective operation often requires continuous optimization of metasurface states in response to changing environments, channel conditions, or sensing objectives\cite{A6}, \cite{A7},\cite{B44}, \cite{A8}, \cite{B50}, \cite{B51}. This typically involves solving large-scale inverse or high-dimensional optimization problems using iterative or data-driven algorithms  \cite{B50}, \cite{A16}, \cite{B52}. In computational imaging and sensing systems, sequential reconfiguration of the metasurface is employed to generate multiple radiation or scattering patterns that must be synchronized with the measurement hardware and incorporated into reconstruction algorithms \cite{A14},\cite{A16},\cite{A11},\cite{A26},\cite{B53}. Such approaches, demonstrated in dynamic metasurface antenna systems, significantly reduce front-end hardware compared to phased arrays but remain limited by active control circuitry and calibration overhead and switching latency \cite{A8, A25, A26, B54, B55}.

While both active and passive metasurfaces have demonstrated impressive performance in wavefront control applications, polarization remains an underutilized degree of freedom in many practical systems\cite{B39},\cite{A7},\cite{A17}. Polarization plays a critical role in electromagnetic sensing and communication, affecting signal detection, channel robustness, and information capacity. Polarization-sensitive metasurfaces have been investigated for polarization-dependent beam routing and multifunctional responses  \cite{B40}, \cite{A19}, \cite{A17}, \cite{B57}. However, most existing designs are optimized for deterministic polarization control rather than for retrieving the polarization state of an unknown incident field  \cite{B39}, \cite{B40}, \cite{A19}.
Recent studies in computational polarimetric imaging have shown that spatio-temporally diverse radiation patterns generated by dynamic metasurface apertures can encode vector electromagnetic information, enabling the recovery of co- and cross-polarized field components from compressed measurements \cite{B37}, \cite{B38}, \cite{A14}, \cite{A20}, \cite{A21},  \cite{B56}, \cite{B58}.
Nevertheless, these systems continue to rely on active tuning networks and multi-port architectures  \cite{A26}, \cite{B55}, \cite{A20}, \cite{A21}. In contrast, fully passive metasurfaces eliminate biasing and control circuitry but generally lack sufficient coding diversity to enable polarization retrieval through computational means \cite{B47}, \cite{A15,A18,B59}. This creates a fundamental gap between hardware-efficient passive structures and flexible but hardware-intensive active metasurface systems  \cite{B49}, \cite{A15}, \cite{A32}, \cite{B36}, \cite{B60}.

To address this gap, we propose a fully passive metasurface aperture designed for polarization-diverse sensing at microwave frequencies. The proposed approach employs a polarization-sensitive meta-atom engineered to exhibit strongly contrasting scattering responses under orthogonal linear polarizations\cite{B37}, \cite{B38}, \cite{A4,A24,A9}. By distributing a randomized ensemble of such anisotropic elements across the aperture, the metasurface generates polarization-dependent, speckle-like scattering patterns  \cite{B47}, \cite{A11}, \cite{B55}, \cite{A15}, \cite{B61}. Sampling the resulting scattered fields at a single output port and applying computational reconstruction techniques enables inference of the incident polarization state without the need for active tuning elements, DC bias networks, mechanical motion, or multi-port receivers\cite{A1}, \cite{A11}, \cite{A26}, \cite{B56}, \cite{A30,A31,B62}.
The proposed architecture offers several advantages over conventional RIS-based or actively tuned polarimetric systems:
\begin{itemize}
    \item Fully passive operation, eliminating power-hungry tuning networks, and simplifying fabrication \cite{B49}, \cite{A32}, \cite{B36}, \cite{B63}.
    \item Polarization diversity through intrinsic anisotropy, combined with engineered spatial randomness that produces rich polarization-dependent coding  \cite{B37}, \cite{B38}, \cite{A16}, \cite{B56}, \cite{B61}.
    \item Compatibility with single-port sensing, aligning with trends in single-pixel imaging and low-cost computational polarimetry.\cite{A14}, \cite{A11}, \cite{A26}, \cite{B55}, \cite{A15}, \cite{B64,B65,B66}.
\end{itemize}
In what follows, we detail the design and simulation of the polarization-sensitive meta-atom (Section II), describe the randomized array architecture and polarization retrieval framework (Section III), present numerical validation and performance analysis (Section IV), and conclude with potential applications in compact polarimetric sensing and imaging (Section V).

\section{Modeling the metasuface configuration}

\subsection{Meta-atom dsign}\label{AA}
\begin{figure}[!t]
    \centering
    \includegraphics[width=0.5\textwidth]{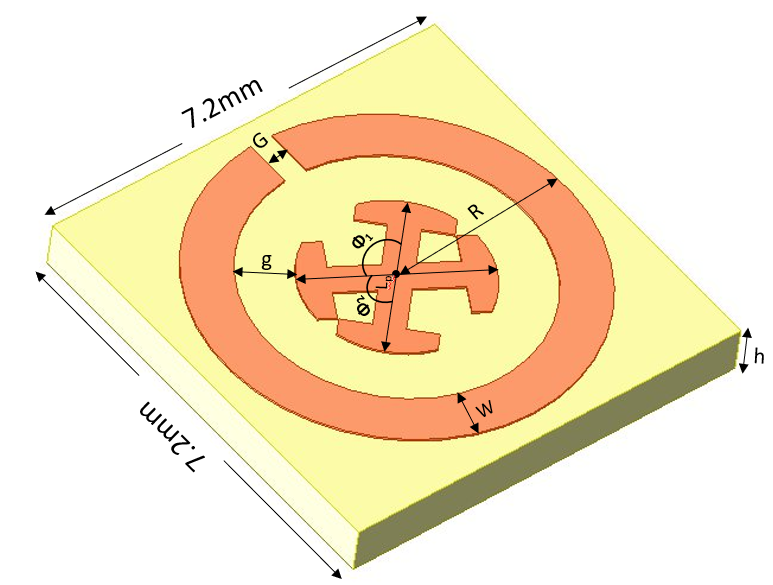}
    \caption{Proposed meta-atom configuration along with the dimensional details. Here, G=$0.4$mm, w=$0.8$mm, R=$3.2$mm, g=$0.9$mm, Lp=$3$mm, and h=$0.813$mm,$\phi_1 = 105^\circ$, $\phi_2 = 75^\circ$.}
    \label{fig 1}
\end{figure}
We implement the proposed Meta-atom on a Rogers RO4003C substrate backed by a copper ground plane. Both the substrate and the ground plane have lateral dimensions of $7.2$ mm × $7.2$ mm, and the substrate thickness is $0.813$ mm. The top metallization layer contains a concentric split-ring resonator (SRR) and an inner hammer-head cross resonator. The SRR has an inner radius of $2.4$ mm and an outer radius of $3.2$ mm, corresponding to a ring width of $0.8$ mm. We introduce a $0.4$-mm slit on the ring to intentionally break structural symmetry, thereby perturbing the surface-current path and modifying the LC resonance behavior. The use of asymmetric and anisotropic resonant elements to enable polarization-dependent electromagnetic responses is consistent with metasurface design principles reported in polarization- and bianisotropy-based metasurfaces \cite{A4,A24,A9}. At the center of the SRR, a hammer-head loaded metallic cross resonator is implemented. Each arm of the cross has an effective length of $1.5$ mm, resulting in a total arm length of $3$ mm including the hammer-head section. The arm width and hammer-head width are each $0.5$ mm, and the hammer-head extension length is $1.4$ mm. We used $0.035$ mm thick conductors throughout the design process.The design intentionally assigns different angular separations to the arms, creating asymmetric angles between adjacent arms on different sides where $\phi_1 = 105^\circ$ and $\phi_2 = 75^\circ$ .The radial spacing between the hammer-head tip and the inner surface of the outer ring, defined as the coupling gap, is $0.9$ mm. 
To verify the polarization-dependent behavior of the Meta-atom, we systematically tuned the geometry of the cross resonator by studying the S-paramater behavior of the unit cell for varying configurations of the cross-arms. To examine the polarization sensitivity of the unit cell we excited the optimized unit cell using normally incident plane waves with orthogonal linear polarizations: along the X and Y directions. Then, the reflection response of the unit cell was evaluated under both polarization states. The comparison of the reflection characteristics under X and Y polarized excitation confirms that the proposed unit cell exhibits strong polarization sensitivity, as illustrated in Fig. $2$. This verification establishes that the electromagnetic response of the meta-atom can be effectively controlled by the polarization of the incident wave. After confirming polarization sensitivity at the unit-cell level, the optimized design was extended to the array configuration.

\begin{figure}[!t]
    \centering
    \begin{minipage}[t]{0.38\columnwidth}
        \centering
        \includegraphics[width=\linewidth]{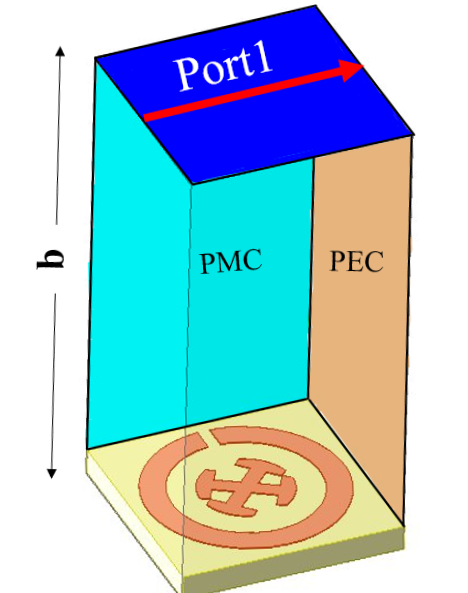}
        \subcaption{}
    \end{minipage}
    \hfill
    \begin{minipage}[t]{0.6\columnwidth}
        \centering
        \includegraphics[width=\linewidth]{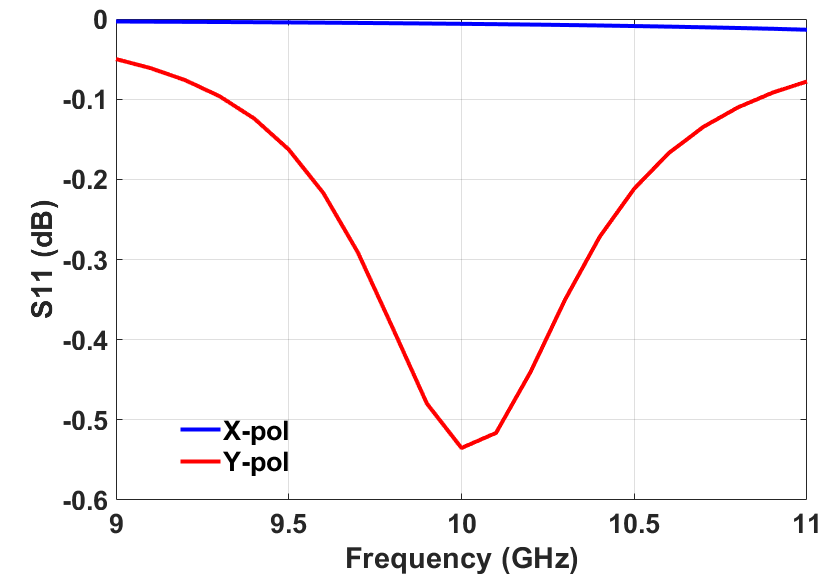}
        \subcaption{}
    \end{minipage}

    \vspace{0.5em}

    \begin{minipage}[t]{0.65\columnwidth}
        \centering
        \includegraphics[width=\linewidth]{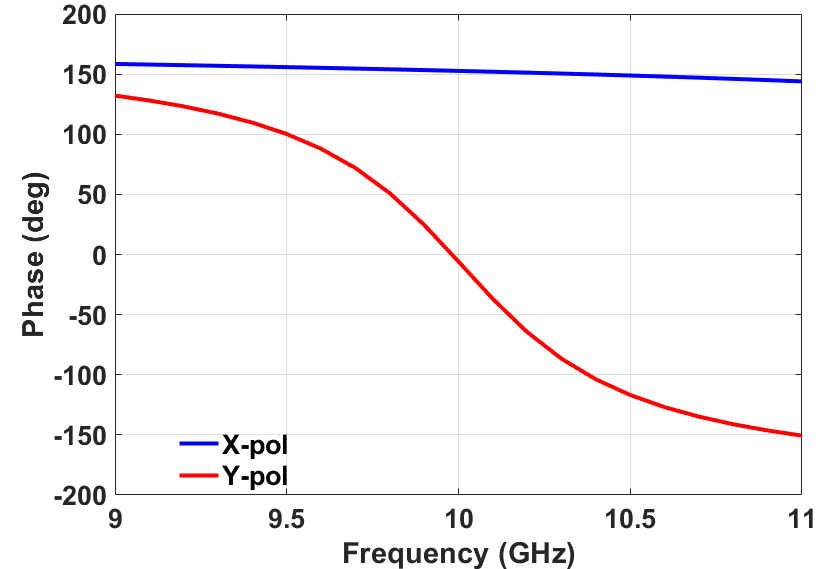}
        \subcaption{}
    \end{minipage}

    \caption{(a) Simulation setup and boundary conditions for a single meta-atom with an airbox of height $b = 40$ mm. (b) $|s_{11}|$ and (c) $\angle{s_{11}}$ under liner polarization.}
    \label{fig 2}
\end{figure}
To simulate the Meta-atom we placed an airbox with periodic boundary conditions on top as shown in Fig $2$(a). The simulation domain measures $7.2$ mm × $7.2$ mm in the planar dimensions and $40$ mm along the z-axis, originating at the ground plane. We selected the domain height to satisfy far-field radiation conditions under normal plane-wave excitation \cite{A11}, \cite{A12},following standard practices in metasurface and dynamic aperture modeling  \cite{B45}, \cite{B52}, \cite{A11}, \cite{A12}. For x-polarized excitation, faces parallel to the x-axis were assigned perfect electric conductor (PEC) boundary conditions, while faces parallel to the y-axis were assigned perfect magnetic conductor (PMC) boundary conditions. A wave port was applied at the top boundary, with the excitation defined using an integration line oriented along the +x direction. For y-polarized excitation, the PEC and PMC boundary assignments were interchanged while maintaining the same wave-port configuration.This periodic excitation framework enables efficient characterization of the polarization-dependent behavior of the proposed meta-atom. Figures~$2$(b) and $2$(c) present the simulated $S_{11}$ magnitude and phase responses under linear polarization excitation, confirming the polarization-sensitive nature of the design and its consistency with established metasurface modeling approaches used for polarization and wavefront control \cite{A4,A24,A9}, as well as computational polarimetric imaging systems   \cite{B37}, \cite{B38}, \cite{A17}, \cite{B56}.
\subsection{Array construction}
After optimizing the Meta-atom, we periodically replicated it along both the X and Y directions to form a $3$ × $3$ metasurface array. The array was generated by duplicating the optimized meta atom eight times and placing each copy adjacent to the original without modifying geometry, spacing, or orientation. This preserves the intrinsic electromagnetic behavior of the Meta-atom while enabling array-level analysis of resonance and polarization response \cite{A8}, \cite{A25} ,consistent with dynamic metasurface antenna and frequency-diverse metasurface array implementations \cite{A42},  \cite{B43}, \cite{B46}, \cite{B48}, \cite{A29}. Following arrangement, a single continuous substrate layer was formed beneath the entire $3$ × $3$ lattice, and the ground plane was extended uniformly across the structure to maintain electromagnetic continuity. To enable computational sensing functionality, We later introduce randomness by varying the orientation or state of individual meta-atoms, thereby generating a diverse set of radiation and scattering patterns. This concept of pattern diversity is foundational to dynamic metasurface antennas and computational sensing systems  \cite{B41},\cite{A8}, \cite{A11},  \cite{A25}, \cite{A26}, \cite{B55}, \cite{B62}.

\begin{figure}[!t]
    \centering
    \includegraphics[width=0.8\linewidth]{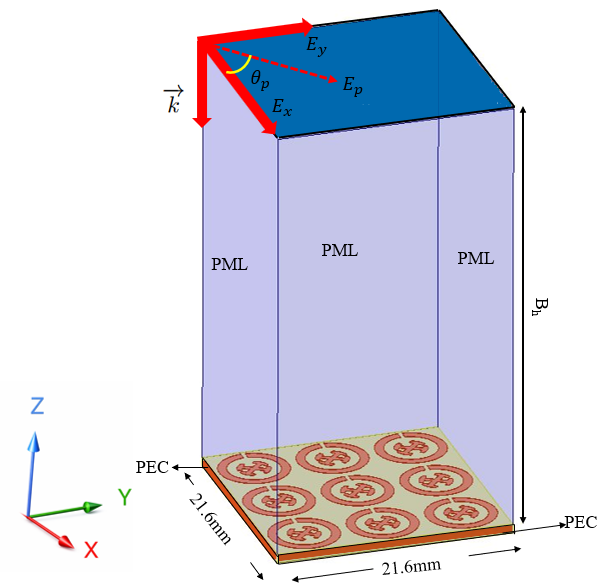}
    \caption{The passive metasurface array and the simulation setup. The polarization-diverse incident wave port is assigned $B_h=65$ mm above the array surface.}
    \label{fig 3}
\end{figure}
The complete array was enclosed within an air region to emulate an unbounded propagation environment. We applied perfectly matched layer (PML) radiation boundaries to all exterior faces of the air region to suppress artificial reflections, following established simulation practices for metasurface-based sensing and imaging systems \cite{B52,A11,A26, B53}, \cite{A12}. An air-box height of $65$ mm was selected to ensure sufficient separation between the metasurface and the radiation boundary, preventing near-field coupling, enabling accurate absorption of outgoing waves \cite{A11}, \cite{A12}, and ensuring plane-wave illumination. It is noteworthy that at $10$ GHz, $65$ mm distance is beyond the far-field boundary of the metasurface dictated by $\frac{2D^{2}}{\lambda}$, where $D$ is the largest dimension of the array i.e., its diagonal, and $\lambda$ is the wavelength at $10$ GHz in free space.  
To evaluate the electric-field variation characteristics with respect to changing polarizations, we first illuminated the array with incident waves of different polarizations. Then, we sampled the field reflected from the metasurface across the array aperture. This is achieved by sensing the reflected field along a line on the top surface of the airbox. Consequently, the sampling line was used to extract tangential electric-field components on the surface for polarization- and frequency-dependent analysis, consistent with field-based measurement strategies employed in computational metasurface sensing systems  \cite{B37}, \cite{B52,A11,A26},  \cite{A12}, \cite{A27}.

\section{Feature extraction and detection protocol}
\subsection{Proposed sensing operation}
We applied a normally incident plane-wave excitation to illuminate the $3 × 3$ metasurface array, with the excitation reference point located at the geometric center of the structure. The incident electric-field vector is defined as $\overrightarrow{E^p} = \langle E_x^p\hat{x},\, E_y^p\hat{y},\, E_z^p\hat{z} \rangle = \langle \cos(\theta_p)\hat{x},\, \sin(\theta_p)\hat{y},\, 0\hat{z} \rangle$ where \( \theta_p \) denotes the polarization angle with respect to the x-axis.

The field propagation direction is set with the propagation vector $\overrightarrow{k} = \langle 0\hat{x}, 0\hat{y}, -1\hat{z} \rangle$ for normal incidence on the surface plane. A linear frequency sweep from $9$ GHz to $11$ GHz with $21$ equally spaced points was performed to characterize the broadband reflection and polarization response of the metasurface. In addition, a parametric sweep of the polarization angle over \(\theta_p \in [-90^\circ, +90^\circ]\) was conducted to assess polarization sensitivity and robustness  \cite{B37}, \cite{B38}, \cite{A11}, \cite{A12}, \cite{A27}. To capture the electric field, we extracted the reflected electric-field from the metasurface in the far-field along the X-axis. We then leveraged the field decomposition technique to acquire  \( X, \)  \( Y, \) and   \( Z \) components of the reflected field separately. In essence, the sensed signal $E_s$ is given by 

\begin{equation}
    \overrightarrow{E^s} = \langle E_x^s\hat{x},\, E_y^s\hat{y},\, E_z^s\hat{z} \rangle 
\end{equation}

Here, A dependency of the $i^{th}$ component of the sensed field on the corresponding incident field component can be tailored through carefully engineering the polarization conversion characteristics of the metasurface array. In this case, the reflective metasurface is passive, allowing only a fixed polarization conversion characteristic. Therefore, incident and reflected field component pairs maintain a relation through the polarization conversion operation on the metasurface. If it is quantified to be encapsulated by a parameter $p$, then     

\begin{equation}
    E_i^s\hat{i} = p E_i^p\hat{i}
\end{equation}

In other words, the polarization of a detected reflected field component retains a strong correlation with the polarization direction of the incident wave. To illustrate, we presented the captured electric field component variations across a few polarization angles of the incident wave on the metasurface at $10$ GHz in Fig. 4. The field component detections for corresponding different incident polarization angle $\theta_p$ are distiguished using three different colors in this figure namely blue, red and green.  
\begin{figure}[!t]
    \centering
    \includegraphics[width=\columnwidth]{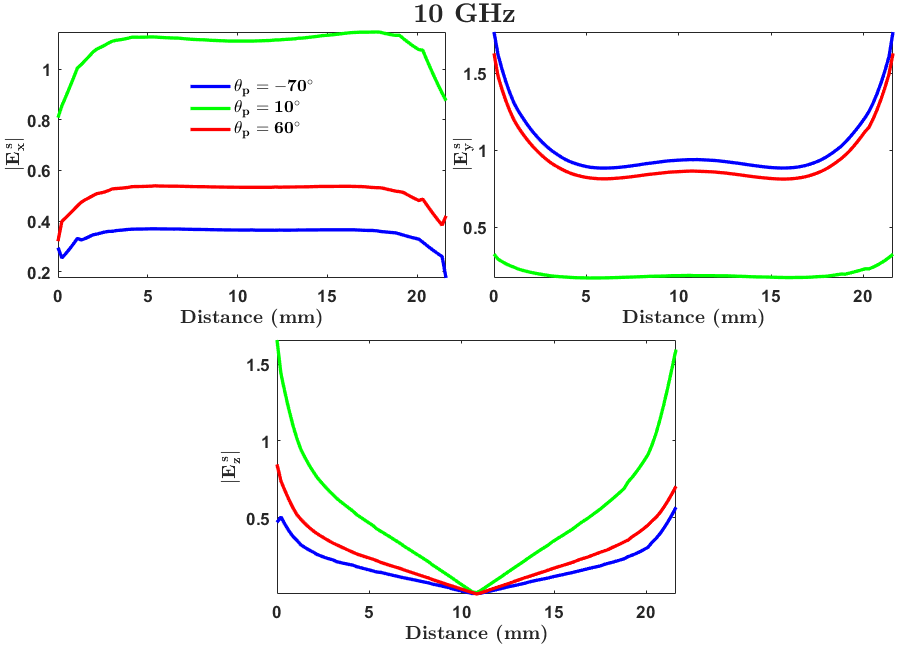}
    \caption{Polarization specific sensing of electric field components at $10$ GHz.}
    \label{fig 4}
\end{figure}

Evidently, all of the three detected electric field components exhibits polarization diverse feature despite the signals were sampled only along the X-axis. The field scattering from the metasurface due to a unipolar excitation scrambles the polarization in all direction. In addition, all the components of the detected signal are dominantly present along the detection line i.e. the X-axis in this case. Therefore, Owing to these phenomena, one can detect the polarization direction of an electromagnetic wave by reflecting the wave using the proposed metasurface configuration towards a linearly polarized antenna with arbitrary orientation. It is worth noting that field extraction along a particular axis or line such as the X-axis is trivial and can be replaced with any arbitrary line. Here, we demonstrated that the polarization of the reflected wave can be effectively detected regardless of the polarization of the detector antenna. In practice, one may employ a linearly polarized antenna with an unknown polarization direction to detect the incoming waves reflected from the metasurface.  

The resulting intensity and phase information from the scattered fields form the measurement vector used by the computational sensing algorithm to infer the incident polarization state. This methodology follows the core principle of computational sensing with dynamic metasurface apertures, where measured scattered fields are used to estimate unknown properties of the incident wave or scene \cite{A1}, \cite{A14}, \cite{B51}, \cite{A11}, \cite{A26}, \cite{B55}, \cite{A12}.
\subsection{The polarization sensitive detection}
\begin{figure}[!t]
    \centering
    \includegraphics[width=\linewidth]{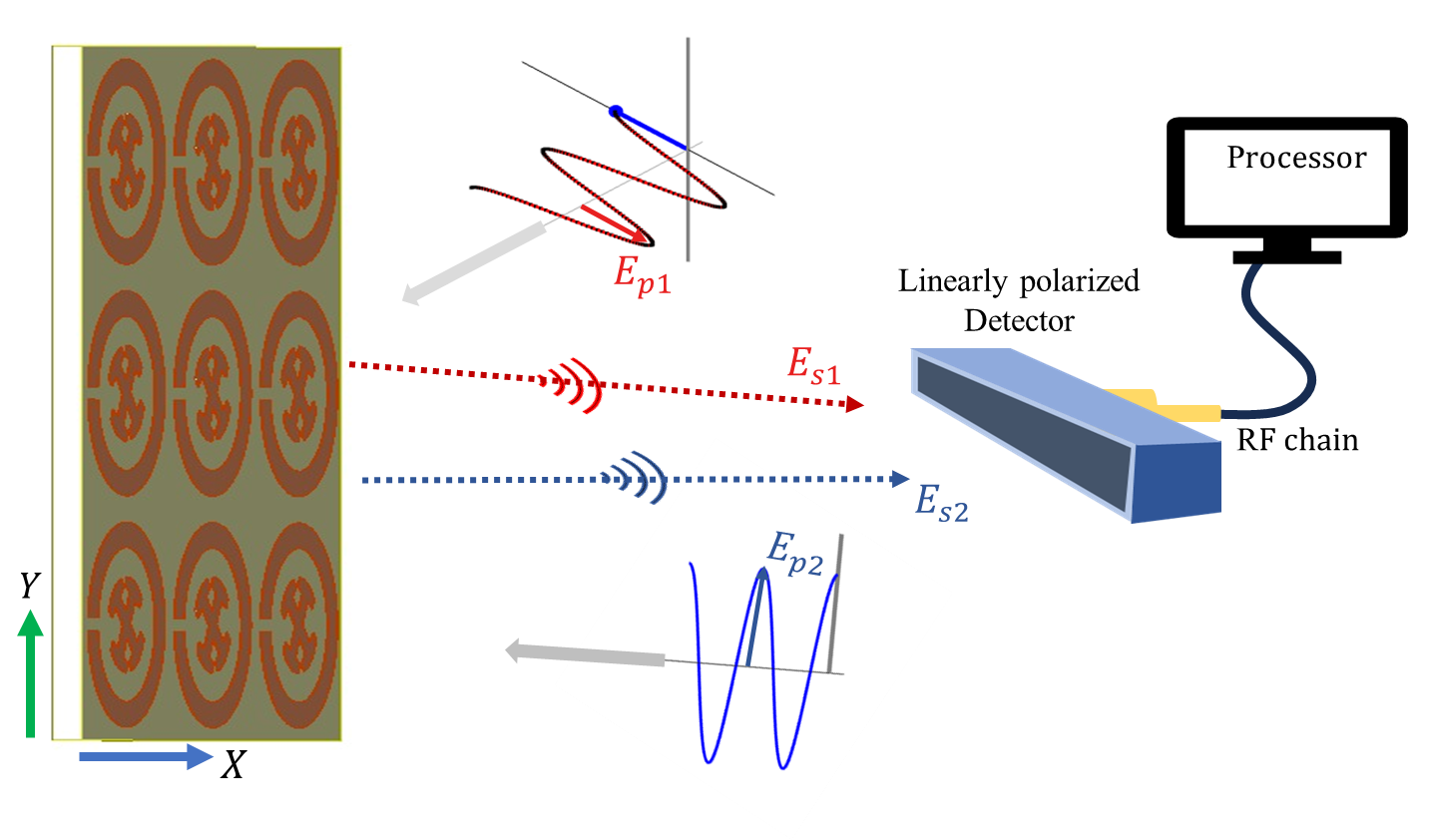}
    \caption{Proposed polarization detection system using a linearly polarized receiving antenna i.e., detector.}
    \label{fig 5}
\end{figure}

The proposed polarization-diverse sensing system operates by exploiting the intrinsic anisotropy and asymmetry of the unitcell comprising a passive metasurface aperture to generate polarization-dependent scattering responses. In essence, at the system level, the metasurface aggregates the polarization-sensitive responses of individual meta-atoms to generate spatially diverse scattering patterns across the aperture. When an electromagnetic wave with an unknown linear polarization is incident on the metasurface, the asymmetric split-ring and hammer-head cross resonator elements induce surface current distributions that are strongly dependent on the orientation of the incident electric-field vector \cite{A4}, \cite{A24}---leading to the coupling between the incident electric field and the resonant elements of the meta-atom to change accordingly. As a result, different polarization states produce distinct reflected field distributions across the metasurface aperture. It should be noted that the metasurface configurations for polarization-sensitive radiation patterns have been explored before for specific applications like polarimetric imaging \cite{A27,B37,B38}. In this work, we propose a passive reflector array instead that is capable of inducing the polarization information into the scattered field. This information can then be detected using a linearly polarized antenna. A linearly polarized receiving antenna samples the scattered electromagnetic fields at a single observation region, enabling a convenient polarization detection system. This simple yet effective polarization detection device is envisioned in Fig. ~\ref{fig 5}. The detector is depicted as a device with a thin rectangular aperture to manifest that any linearly polarized antenna with the same operation frequency band as the metasurface can be used for desired detection.    

In the detection setup of Fig. ~\ref{fig 5}, the metsaurface array is placed on the XY-plane. Thus, normal incident and reflection directions are a=long the -Z and +Z axes, respectively. To illustrate the proposed detection operation, we used two incidence-reflection pairs denoted with numbers $1$ and $2$. As each of the two incoming waves $E_{p1}$ and $E_{p2}$ of different polarizations illuminates the metasurface array, the metasurface scatters waves with distinctive signatures for each incident wave. These polarization-dependent scattering signatures manifest as variations in the amplitude and phase of the reflected fields, forming spatially diverse radiation patterns without requiring active tuning elements or reconfigurable circuitry. The retained polarization-sensitive magnitude and phase information on the detection line (see Fig. 3) for incident waves of a wide range of polarizations between $[-90^\circ,90^\circ]$ is provided in Fig. 6. Clearly, the detected signal carries distinguishable polarization-specific signatures both in magnitude and in phase. 
\begin{figure}[!t]
    \centering
    \includegraphics[width=\linewidth]{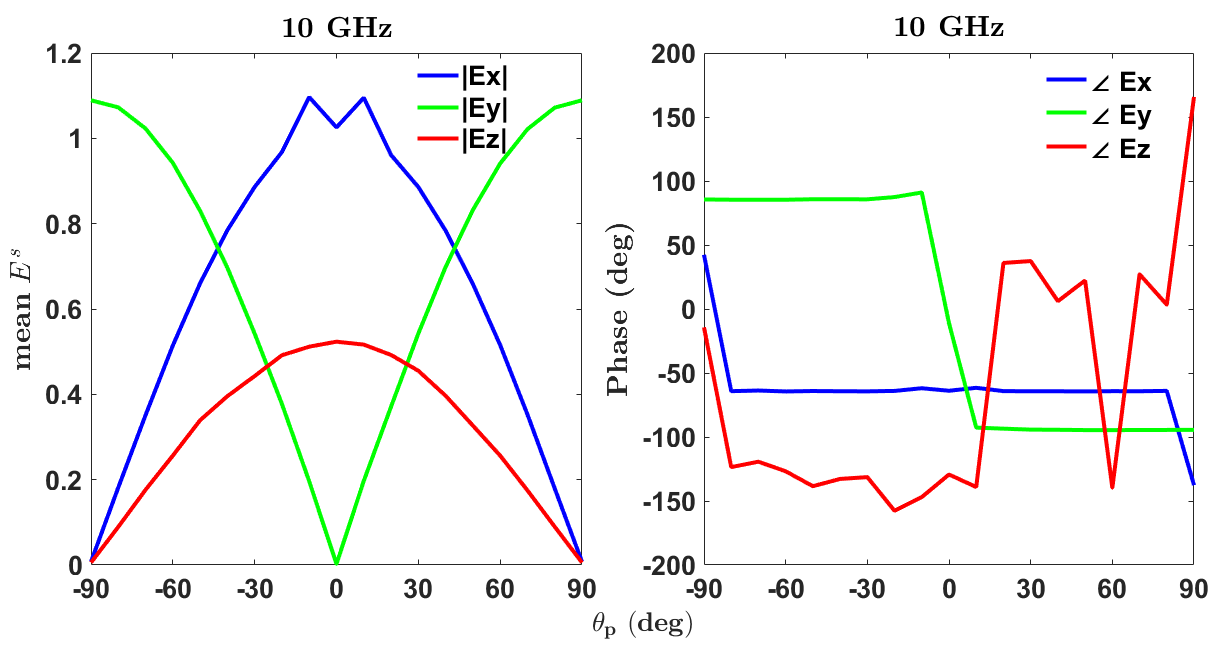}
    \caption{Mean electric-field (left: magnitude, right: phase) reflection from the metasurface for each incoming polarization angle at 10 GHz.}
    \label{fig 6}
\end{figure} 
The detection system captures the scattered electromagnetic fields at a single output port or sampled observation region, providing a compact measurement interface. Rather than directly measuring polarization through conventional polarimetric hardware, the proposed system adopts a computational sensing paradigm. As shown in the example setup in Fig. 5, a single RF connector (or an RF chain) can be used to carry the signal into a processing unit. In the processor, the complex scattered-field data, including intensity and phase information, are treated as a sensing vector whose structure encodes the polarization content of the incident wave. By applying numerical reconstruction or inference algorithms to these sensed data, the incident polarization state can be estimated from the measured scattering signatures. This approach enables polarization-diverse sensing using a fully passive metasurface aperture, eliminating the need for biasing networks, multiport receivers, or mechanically reconfigurable components  \cite{A1}, \cite{A25},  \cite{A32}, \cite{B36} , \cite{A30},  \cite{B66}, \cite{A28}.

It is worth noting from Fig. ~\ref{fig 6} that the polarization-specific signatures are strongly noticeable in the magnitude or intensity of the detected signal. Therefore, one can perform the desired polarization detection operation using only the intensity data. Thus, the proposed detection system can facilitate utilization of a single coaxial cable instead of complex RF chains (see Fig. ~\ref{fig 5}) to carry sensed data from the antenna to the processing unit. As a result, the proposed system enables intensity-only detection without requiring complex phase detection circuitry. Nevertheless, in case the phase information is needed for some specific application, it is triavial to attach a phase detector circuitry to the system shown in Fig. ~\ref{fig 5}. The polarization sensitive phase data presented in Fig. ~\ref{fig 6} indicates that the polarization can be detected using only phase data as well. 

However, for clarity and conciseness, we will employ only intensity data in the remainder of this paper. The efficacy of the intensity data in retaining the polarization state, as demonstrated in Fig. ~\ref{fig 6}, is convincing for a single frequency. While polarization at a single frequency may serve the purpose of mere polarization sensitivity, the basis of the proposed operation presented here aims to serve beyond just polarization detection. In this developmental endeavor, we have devised a simplified system geared towards the advancement of a plethora of polarimetric operations. 
\subsection{Inclusion of frequency diversity}
As the proposed detection scheme retains distinctive polarization specific signatures, it can be utilized to induce polarization diversity to enhance information in compressed data. In fact, previous works have demonstrated information enhancement through multiplexing surface reactance diverse \cite{B67, A28} and frequency diverse \cite{A12, B68} data. As surface reactance is a fixed entity in the case of a passive metasurface, frequency diversity can be induced by spatially varying diverse resonant elements with fixed architectures \cite{B68}. This is not entirely the case for our passive metasurface structure, as the metasurface is homogenously filled with the same passive elements. Therefore, based on the polarization sensitive of the unitcell as shown in Fig. ~\ref{fig 6}, we relied on the polarization diversity to enhance the information in the sensed data. 

However, polarization diversity alone may not be sufficient to serve the purposes of many applications. For instance, according to the symmetric trends visible across the range of polarization angles in Fig. ~\ref{fig 6}, the detection of polarization angles symmetric about zero (e.g. $\pm30^\circ$, $\pm45^\circ$, etc.) with only polarization diverse data is ambiguous. This issue can potentially be addressed with the inclusion of frequency diversity along with the polarization diverse data. This stems from the fact that the responses of the unit cell maintain decent variations over a bandwidth around the resonant frequency, as shown in Fig. ~\ref{fig 2}b-c. We have empirically found that the variations in the reflections from the unitcell retained over $9-11$ GHz are suitable for our demonstration. Generally, this operational bandwidth can be adjusted to meet specific operational criteria. 

\begin{figure}[!t]
    \centering
    \includegraphics[width=\linewidth]{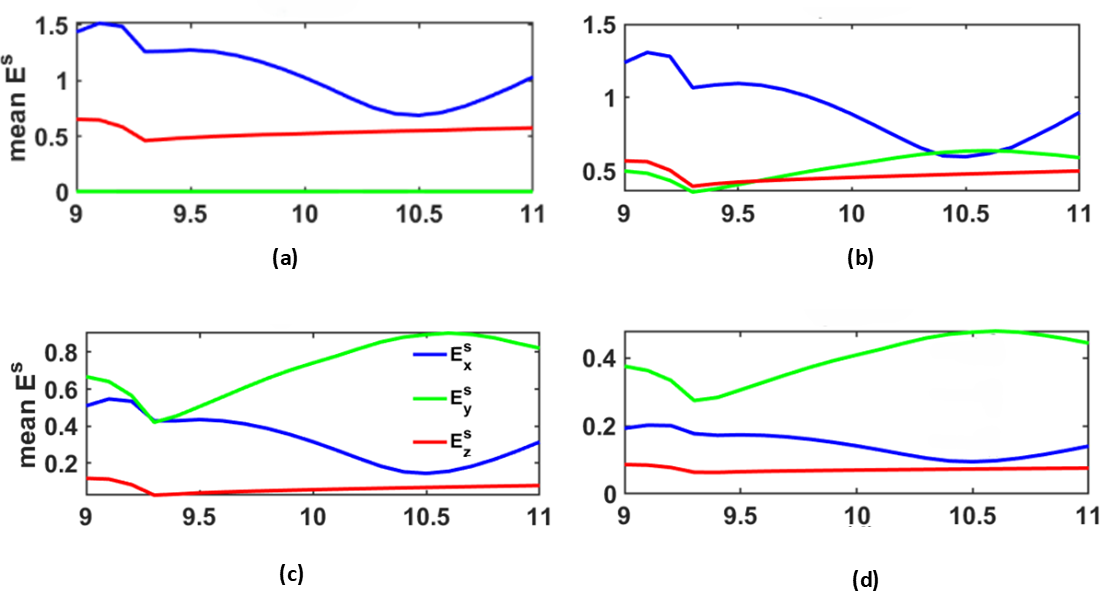}
    \caption{Sensed mean electric-field reflection vs. frequency for (a)$\theta_p=0 ^\circ$ (b)$\theta_p=30 ^\circ$ (c)$\theta_p=60 ^\circ$ (d)$\theta_p=70 ^\circ$.}
    \label{fig 7}
\end{figure}

When implemented on the array, the reflections from all the elements are coupled to create multiplexed field distributions in the far-field. The frequency diverse field distributions of $E_x, E_y$, and $E_z$ components for different incident polarization angles are depicted in Fig. ~\ref{fig 7}a)-d). The impact of exciting the metasurface with a frequency diverse each linearly polarized wave is strongly noticeable in all the plots. Thus, the polarization diversity combined with frequency diversity has the potential to deliver improved detection performance.        

\begin{figure}[!t]
    \centering
    \includegraphics[width=0.8\linewidth]{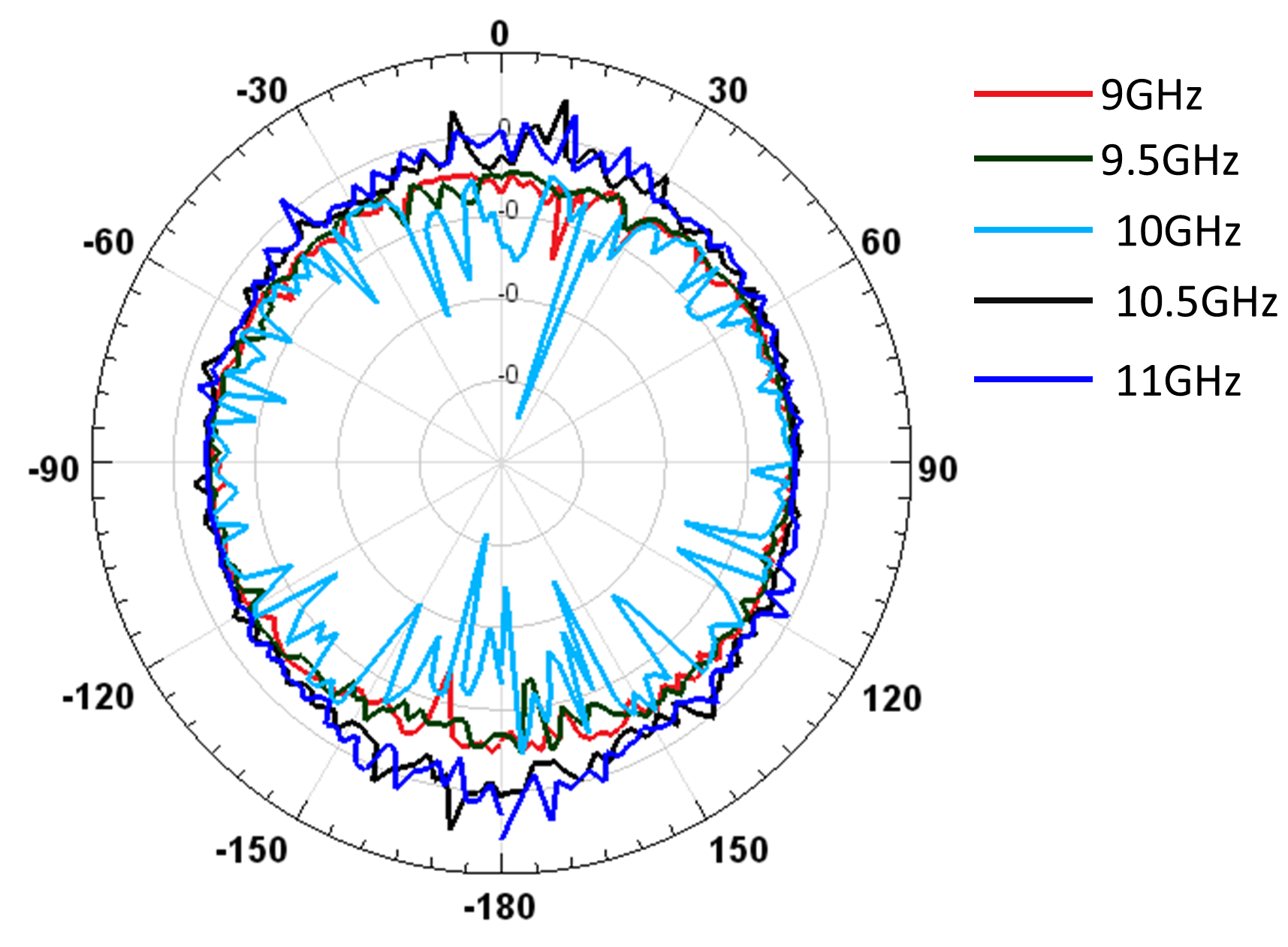}
    \caption{Reflection pattern of the metasurface for $\theta_p = 0^{\circ}$ at a set of frequency points.}
    \label{fig 8}
\end{figure}
\begin{figure}[!t]
    \centering
    \includegraphics[width=0.75\columnwidth]{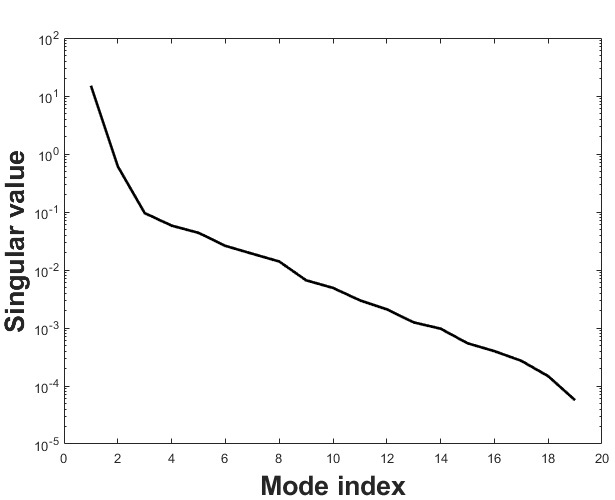}
    \caption{Singular value decomposition of the captured data}
    \label{fig 9}
\end{figure}

To further investigate the impact of frequency diverse incident waves, we observed the reflection patterns of the metasurface at individual frequency points. Fig. ~\ref{fig 8} illustrates representative radiation characteristics of the metasurface under illumination at different frequencies. The distinctiveness of these patterns implies negligible, if any, correlation among the scattered fields corresponding to different frequency excitations. This observation supports the hypothesis that measurements acquired at additional frequencies contribute genuinely new information about the incident signal. To justify any weakness in this hypothesis, even after empirically identifying uncorrelated frequency diversity, we statistically examined the sensed data at al the frequency points. In particular, we decomposed the data sequences into their singular values to quantify the degree of novelty being added with an increasing number of frequency points in the captured data. The singular value decomposition (SVD) of the sensed data is presented in Fig. ~\ref{fig 9}. The gradually decreasing slope of the SVD highlights that the frequency specific data are weakly correlated and that the information in the sensed data is being enhanced with the inclusion of new frequency points.  
\section{Demonstration of proposed detection operation}
Because we combined both the polarization and frequency diversity into our combined data for versatile and practical detection, the one-to-on correlation between the sensed field intensity and the incoming polarization signal is not retained. Consequently, we resorted to leveraging an estimation algorithm to resolve the desired polarization information. To formulate the estimation problem, we began by constructing a refernce matrix $\mathbf{\Phi_f}$ containing sensed data for $19$ polarization angles between $[-90^\circ,90^\circ]$ at $21$ aforemntioned frequency points. Thus, $\mathbf{\Phi_f}$ is a $21\times19$ matrix. If the metasurface is illuminated with a wave with unknown polarization $\mathbf{\theta_p}$, then the estimation problem can be expressed as follows.

\begin{equation}
    \mathbf{s} = \mathbf{\Phi_f} \mathbf{\theta_p}
\end{equation}

\begin{figure}[!t]
    \centering
    \includegraphics[width=0.75\columnwidth]{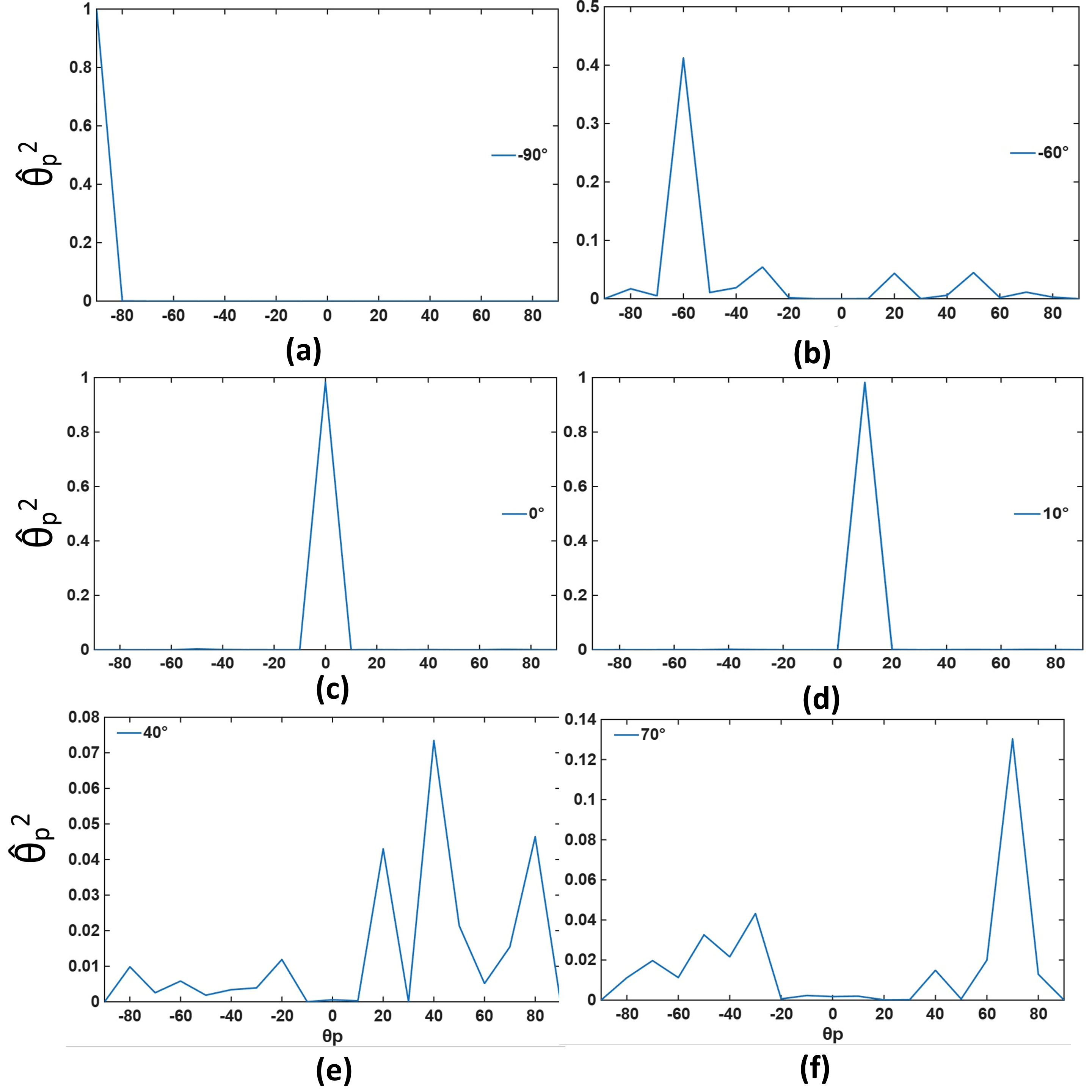}
    \caption{The estimated $\hat{\mathbf{\theta_p}}$ for a set of incident polarization angles.}
    \label{fig 10}
\end{figure}

Where $s$ is the sensed data at the detector. To effectively solve this inverse estimation problem, we used an iterative solver. While a number of suitable iterative solvers are readily available, we empirically found that the \textit{least squares} solver is just the right tool for the estimation operation. A thorough statistical analysis to rank the best algorithm to solve the estimation problem is beyond the scope of this paper and is left for future work. Here, we computationally demonstrate the feasibility of retaining polarization information in the incoming electromagnetic wave using the proposed metasurface. This polarization information is retained by the estimation of $\mathbf{\theta_p}$, which is a vector comprising $19$ elements for $19$ polarization angles. Ideally, $\mathbf{\theta_p}$ would have a peak only at the index corresponding to the incident polarization angle. In reality, however, the reflected field undergoes inter-element coupling. In addition, the array aperture had to be kept comparatively small for the ease of full-wave simulation, potentially resulting in edge effects and undesired perturbations. Moreover, we spatially detected the reflected signal across a line and then took the mean of the spatial data. In the process, any undesired perturbation across the array aperture can disturb the detected signal. Consequently, the estimator of $\mathbf{\theta_p}$, denoted here as $\hat{\mathbf{\theta_p}}$, may exhibit some redundant peaks. However, as illustrated in Fig. ~\ref{fig 10}, the peaks corresponding to the incoming polarization angles are always dominant. To further enhance the detectability, one may utilize $n{th}$ power of $\hat{\mathbf{\theta_p}}$. Here, we presented the estimation capability with $n=2$. Clearly, the proposed detection system can successfully capture the polarization information of a radio wave.     

\begin{figure}[!t]
    \centering
    \includegraphics[width=0.75\columnwidth]{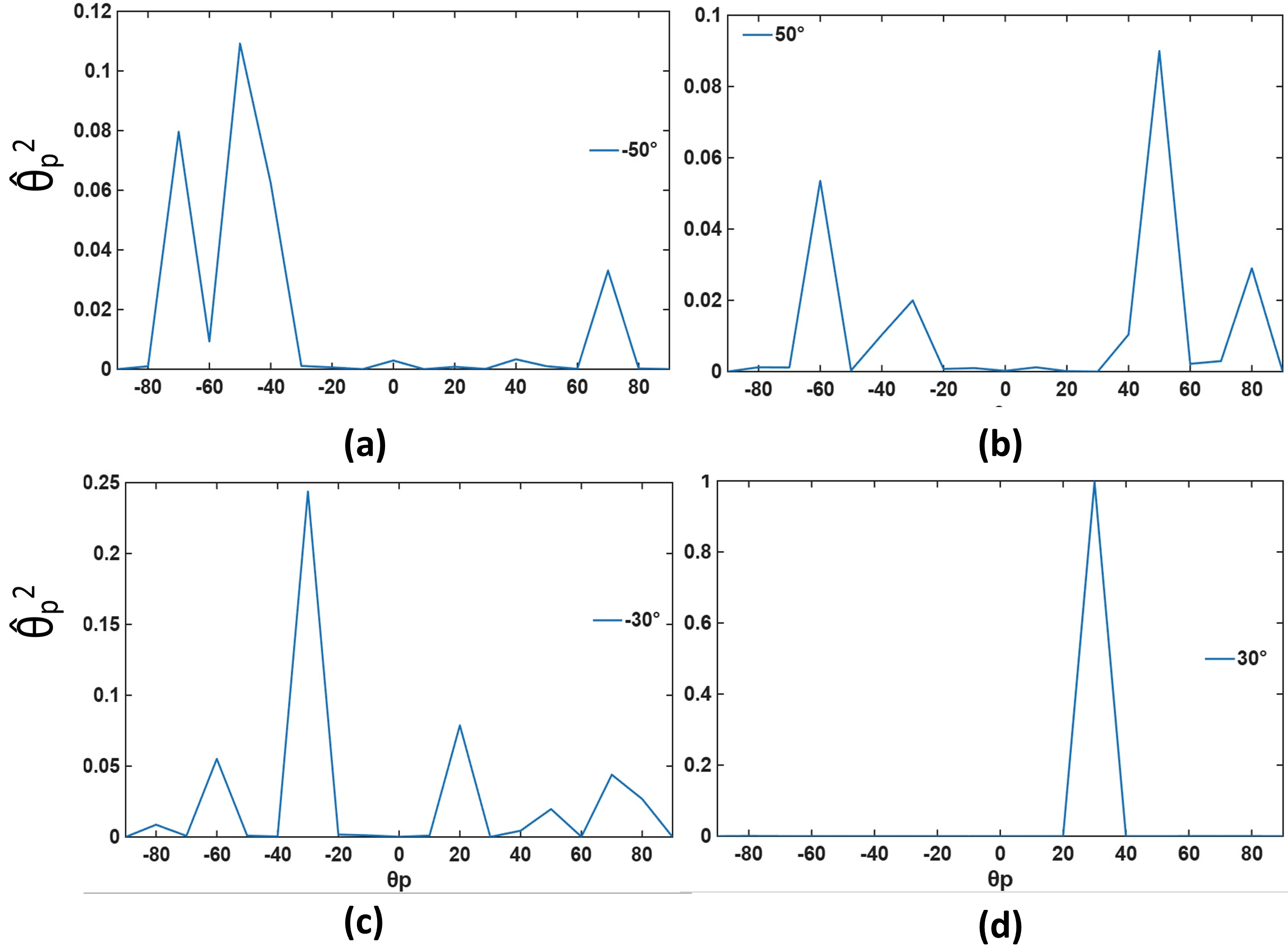
    }
    \caption{The estimated $\hat{\theta_p}$ for a set of incident polarization angles symmetric about zero.}
    \label{fig 11}
\end{figure}

A crucial aspect of polarization sensitivity is the detection of polarization angles symmetric about zero, which could not be achieved through only polarization-diverse detection shown in Fig. ~\ref{fig 6} with successive discussion. With the introduction of the estimation algorithm, we could successfully discern two signed polarization angles with opposite polarity. A few samples of such estimation performance are depicted in Fig. ~\ref{fig 11}, evincing the fact that by incorporating frequency diversity with the polarization diverse sensing more robust and complex detection capability can be achieved.    

\section{Discussion and conclusion}
This work has presented a fully passive metasurface aperture for polarization-diverse detection at microwave frequencies, built around an anisotropic meta-atom that exhibits strong polarization sensitivity and extended to an array whose scattered fields encode polarization information over frequency. Through full-wave simulations, it was shown that the metasurface generates distinct intensity and phase signatures for different incident polarization angles, and that frequency diversity in the 9–11 GHz range significantly enhances the information content of the sensed data. A computational estimation framework successfully recovers the incident polarization angle, including signed angles symmetric about zero. 

The proposed architecture eliminates the need for DC biasing, tuning networks, or mechanical reconfiguration, unlike conventional reconfigurable intelligent surfaces and actively controlled polarimetric apertures. As a result, it provides a scalable and fabrication-friendly platform for polarization-sensitive sensing with minimal RF front-end complexity, leveraging computational processing to extract vector field information from compressed scalar measurements. These characteristics position the presented metasurface as a strong candidate for integration into compact polarimetric sensors and computational imaging systems in applications where size, power, and cost are tightly constrained.

In this manuscript, we present a developmental study aimed at realizing an autonomous detection device for integration into a broad range of polarimetric applications. Building on this foundation, several key research directions will be pursued in future work. First, we aim to enhance detection performance by scaling the metasurface to a larger array, which was not feasible in the present study due to computational limitations. In addition, we plan to investigate randomized metasurface configurations as an alternative to the current homogeneous array, with the goal of increasing coding diversity and improving robustness against noise and fabrication tolerances. As an immediate next step, we will extend the estimation framework by incorporating advanced reconstruction techniques, including sparse recovery, Bayesian inference, and deep neural networks trained on both simulated and experimental datasets. These approaches are expected to improve estimation accuracy, mitigate sensitivity to secondary peaks, and enable real-time polarization tracking. 

Most importantly, future efforts will focus on fabricating the proposed metasurface and conducting anechoic-chamber measurements to experimentally validate the scheme under practical hardware impairments, such as connector losses, mutual coupling, and frequency-dependent dispersion, while developing lightweight calibration protocols. Beyond the scope of this work, the proposed approach may also stimulate further research into embedding passive, polarization-diverse metasurfaces within MIMO- or RIS-assisted communication systems and radar front ends, where joint exploitation of spatial, frequency, and polarization diversity could enable improved channel estimation, target classification, and environment-aware beamforming.

\bibliographystyle{IEEEtran}
\bibliography{new}

\end{document}